\documentstyle[prl,aps,epsf]{revtex}
\begin{document}
\twocolumn[\hsize\textwidth\columnwidth\hsize\csname
@twocolumnfalse\endcsname

\draft
\title{Dynamics of the 1D Heisenberg model and optical
 absorption of spinons in cuprate antiferromagnetic chains.}
\date{\today }
\author{J. Lorenzana}
\address{Centro At\'{o}mico Bariloche, 8400 S. C. de Bariloche, Argentina. }
\author{R. Eder}
\address{Laboratory of Applied and Solid State Physics, Materials Science Centre,\\
University of Groningen,Nijenborgh 4, 9747 AG Groningen, The Netherlands}
\maketitle

\begin{abstract}
\widetext
We use numerical and analytical results to construct a simple ansatz for the energy 
dynamical correlation function of the 1D antiferromagnetic Heisenberg model.
 This is applied to compute the phonon assisted absorption spectra of magnetic 
excitations (spinons) in quasi-one dimensional spin 1/2 insulators
and show to reproduce very well recent infrared measurements in Sr$_2$CuO$_3$.
\end{abstract}

\vskip2pc] \narrowtext

After the solution by Bethe and many years of intense research, the spin 1/2
Heisenberg chain is one of the most studied many-body problems. In spite of this
little is known on its dynamical properties. This is because the extreme
complexity of Bethe's wave functions makes it too hard to compute dynamical
correlations. Much of the progress came from numerical studies which
inspired an ansatz for the dynamical structure factor\cite{mul79,mul81}, a
recent analytical evaluation of the exact two-spinon contribution\cite{bou96}
and low energy bosonization studies\cite{hal82,hal82a,cro79}.

In this letter we use numerical and known analytical results to obtain
a  simple ansatz for the
energy dynamical correlation function defined below. This correlation
function is important when considering the problem of coupling with phonons%
\cite{cro79} and in optical properties\cite{lor95,lor95b}. As an application
we compute the line shape for optical absorption of magnetic excitations
assisted by phonons\cite{lor95,lor95b} and show how spinon physics appears
naturally in the spectra. This is shown to reproduce very well recent
measurements of the line shape by Tokura {\it et al}. \cite{tok96,suz96} in
quasi-one dimensional cuprates (See Fig. \ref{adw}).

We will consider the 1-dimensional (1D) Hamiltonian,
\begin{equation}
H=\sum_i^NJB_i  \label{h}
\end{equation}
where 
\begin{equation}
B_i=S_i^xS_{i+1}^x+S_i^yS_{i+1}^y+\Delta S_i^zS_{i+1}^z  \label{b}
\end{equation}
is the bond-energy operator, $\Delta $ measures the anisotropy and we take
periodic boundary conditions. We will consider $\Delta =0$ (the XY model)
and $\Delta =1$ (the antiferromagnetic Heisenberg model).

The energy correlation function is defined as \cite{cro79}, $G(p,\omega
)=\langle \!\langle \delta B_p;\delta B_{-p}\rangle \!\rangle (\omega )$ (in
Zubarev's notation\cite{zub60}) where $\delta B_p$ is the Fourier transform
of $\delta B_i\equiv B_i-\langle B_i\rangle .$

To warm-up we will compute $G(\omega ,p)$ for $\Delta =0$. This is trivially
done by using a Jordan-Wigner transformation\cite{sol79} to map the problem
to a noninteracting 1D free fermion gas. Excitations are free fermions and $%
G(p,\omega )$ becomes a Lindhard like response describing density like
fluctuations. We get for the imaginary part,

\begin{equation}
-2%
\mathop{\rm Im}
G_{XY}(p,\omega >0)=D_{XY}(p,\omega )M_{XY}(p,\omega )  \label{imgxy}
\end{equation}
where following the approach of Ref. \cite{mul81} we defined a density of
states $D(p,\omega )$ and a matrix element $M(p,\omega )$ . 
\begin{equation}
D_{XY}(p,\omega )=\frac N{2\pi }\frac{\theta (\omega -\omega
_1^{XY}(p))\theta (\omega _2^{XY}(p)-\omega )}{\sqrt{\left[ \omega
_2^{XY}(p)\right] ^2-\omega ^2}}  \label{dosxy}
\end{equation}
with $\omega _1^{XY}(p)=J\sin (p)$, $\omega _2^{XY}(p)=2J\sin (p/2)$ and 
\begin{equation}
M_{XY}(p,\omega )=2\frac{\left[ \omega _2^{XY}(p)\right] ^2-\omega ^2}{%
N^2\left[ \omega _2^{XY}(p)\right] ^2}
\end{equation}

Two-fold degenerate states are counted once in Eq.~(\ref{dosxy}) and this is
taken into account by the factor of $2$ in the matrix element (See Ref. \cite
{mul81}).

Notice that $%
\mathop{\rm Im}
G_{XY}(p,\omega )$ is only different from zero between $\omega _1^{XY}$ and $%
\omega _2^{XY}$ . For a finite system the spectrum is given by a
two-parameter
 set of discrete states with energy and pole strength given for $%
p>0 $ by, 
\begin{equation}
\begin{tabular}{l}
$\omega _{XY}(p,q)=\omega _2^{XY}(p)\cos (q-p/2)$ \\ 
$M_{XY}(p,q)=\frac 2{N^2}\sin ^2(q-p/2)$%
\end{tabular}
\label{sdpq}
\end{equation}
where $0<q\leq p/2.$ Eq. (\ref{sdpq}) corresponds to particle-hole
excitations around the $-\frac \pi 2$ Fermi point.

When we switch to the isotropic limit ($\Delta =1$) a simple
analytic way of computing $G(p,\omega)$ is not available. 
Next we will follow  Ref.~\cite{mul81} and use numerical
results in finite chains, known low energy bosonization results and 
Bethe ansatz results to construct a phenomenological ansatz for 
$\mathop{\rm Im} G(p,\omega )$. 

In the isotropic limit there is no upper bound
for the support of $%
\mathop{\rm Im}
 G(p,\omega )$ however the dominant contribution to $%
\mathop{\rm Im}
G(p,\omega )$ comes from singlet states between $\omega _1(p)$ (the des
Cloizeaux-Pearson\cite{clo62} dispersion relation) and $\omega _2(p)$ which
are obtained from $\omega _1^{XY}(p)$ and $\omega _2^{XY}(p)$ by changing $J$
according to 
\begin{equation}
J_{XY}\rightarrow J\pi /2  \label{xy2h}
\end{equation}
and we attached an $XY$ subindex to avoid confusion latter. This is
analogous to the scenario found for the dynamical structure factor\cite
{mul81} and corroborated by a recent analytical evaluation which retains
only two-spinon excited states\cite{bou96}. We illustrate the analogous
behavior for the energy correlation function for a finite system in Fig. \ref
{mdwp}.

\begin{figure}[tbp]
\epsfverbosetrue
\epsfxsize=6cm
$$
\epsfbox[150 450 562 743]{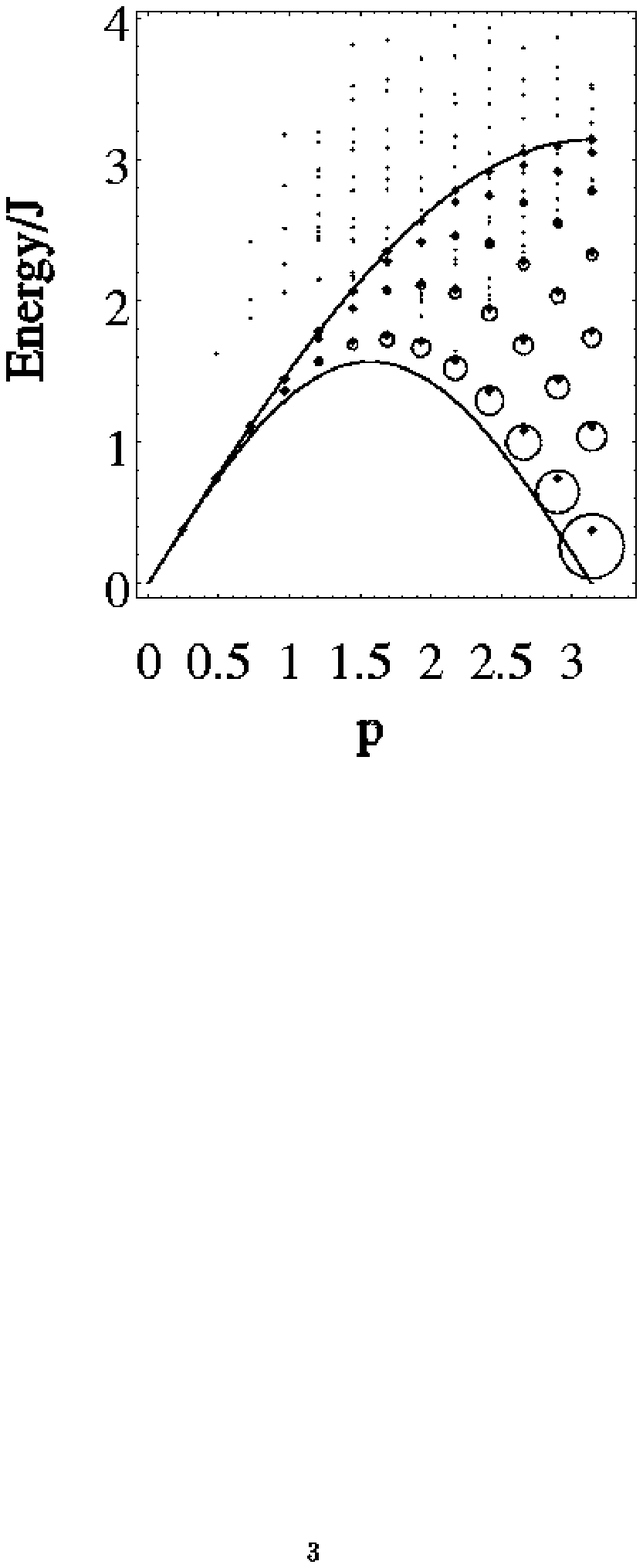}
$$
\caption{Exact diagonalization results for $%
\mathop{\rm Im}
G(p,\omega )$ in the $\omega -p$ plane for a 26 site Heisenberg ring.
Excitations are given by the position of the circles and matrix elements by
the radius. The dots are the XY energies with the renormalization of Eq. (%
\ref{xy2h}). The lines are $\omega _1(p)$ (lower) and $\omega _2(p)$
(upper).}
\label{mdwp}
\end{figure}
In the thermodynamic limit these states form a two-parameter continuum
degenerate with the triplet states that dominate the dynamical correlation
function and have a density of states, $D(p,\omega
)=D_{XY}^{J_{XY}\rightarrow J\pi /2}(p,\omega )$ \cite{mul81}.

To construct an ansatz for the energy correlation function of the Heisenberg
model we put

\begin{equation}
-2%
\mathop{\rm Im}
G(p,\omega >0)=D(p,\omega )M(p,\omega ).  \label{img}
\end{equation}
Our ansatz for $M(p,\omega )$ is given by, 
\begin{equation}
M(p,\omega )=M_{XY}^{J_{XY}\rightarrow J\pi /2}(p,\omega )(A\frac{\sqrt{J}}{%
\sqrt{\omega }}+B\frac J{\sqrt{\omega ^2-\omega _1(p)^2}})  \label{igdw}
\end{equation}

\begin{figure}[tbp]
\epsfxsize=8cm
$$
\epsfbox[150 519 562 743]{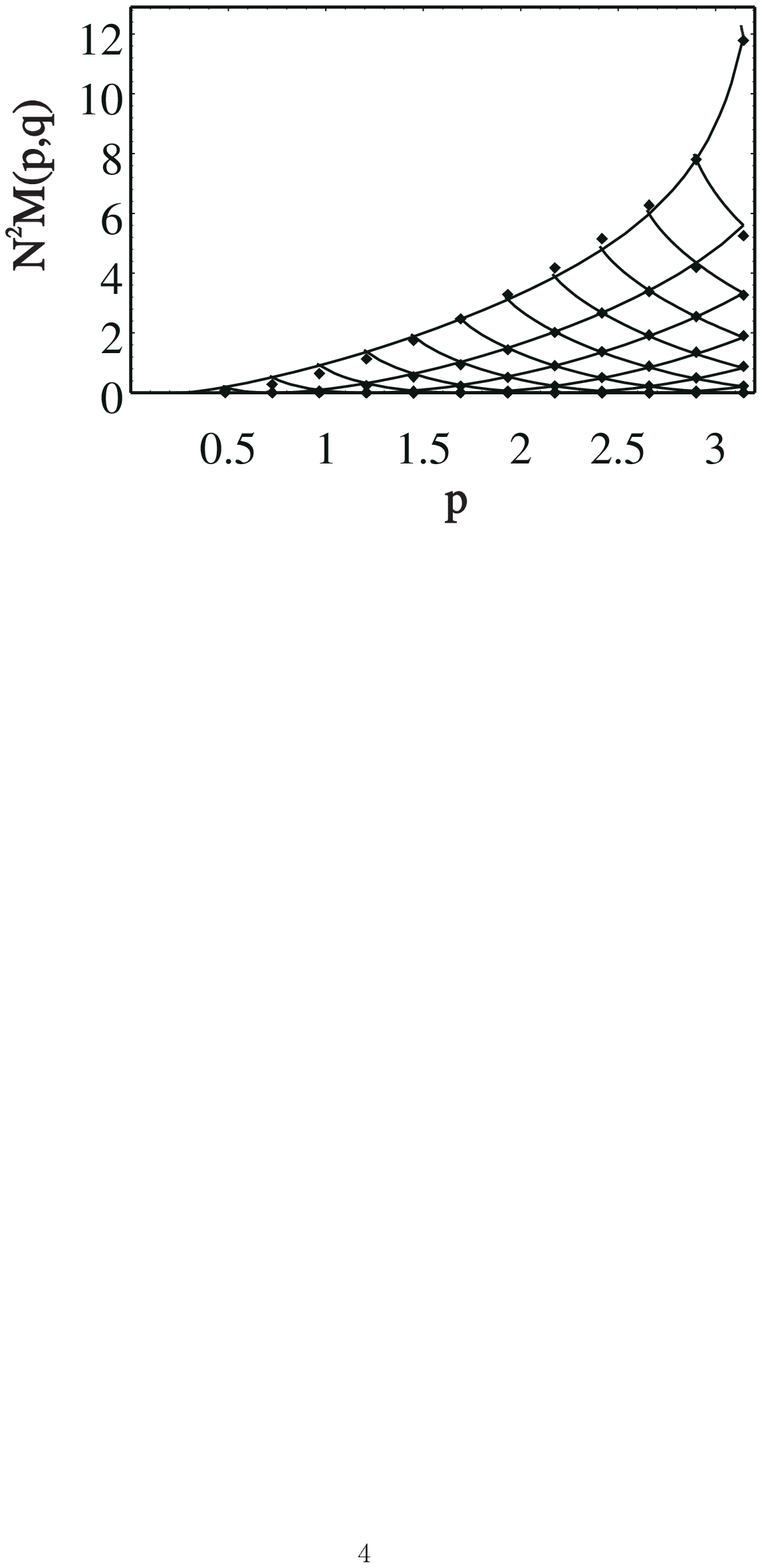}
$$
\caption{The dots are the matrix elements (times $N^2$) vs. $p$ from exact 
diagonalization of a 26 site Heisenberg ring. We had plotted the fit of
$N^2 M(p,q)$ (the ansatz) as continuous functions of $p$ for each allowed 
$q$ in the 26-site chain. The intersection of the continuous curves
gives the actual values for the finite chain.}
\label{mdp}
\end{figure}

For a finite system we use Eq 's (\ref{sdpq}) to construct $M(p,q)$ in the
same manner. In Fig. \ref{mdp} we plot the matrix elements of Fig. \ref{mdwp}
as a function of $p$ together with the best fit of $M(p,q)$ from the ansatz.
The agreement is very good for all branches with significant spectral
weight. 

It is interesting that one can fit all branches except the
upper one with $A\simeq 3$ and $B=0.$ The upper branch corresponds to the
states closer to $\omega _1(p)$ where the square root becomes important.
i.e. the dominant correction to the XY matrix elements in most of the $%
(p,\omega )$ plane is given by the $p$ independent $1/\sqrt{\omega }$ factor.

To determine $A$ and $B$ for an infinite system we have made a finite size
scaling study and used the sum rule $\sum_p\int d\omega (-2)%
\mathop{\rm Im}
G(p,\omega )=\ln 2-(\ln 2)^2$ to get $A=2.4$, $B=0.6$.

Bosonization studies predict\cite{hal82,hal82a,cro79} a divergence for
$-\mathop{\rm Im}
G(p=\pi,\omega )$
like $\omega ^{-1}$ which becomes $(\omega -\omega _1(p))^{-1/2}$
close but not at $p=\pi .$ The term proportional to $B$ in $M(p,\omega )$
ensures this behavior and is analogous to the form introduced by Muller {\it %
et al}. \cite{mul81} for the dynamic structure factor . As mentioned before
its importance here is quantitatively minor in most of the $(p,\omega )$
plane whereas in the case of the dynamic structure factor it was the
dominant contribution.

We next apply our results to the computation of the line shape for   
phonon assisted optical absorption of magnetic excitations for a 1D
spin 1/2 isotropic Heisenberg chain. The absorption coefficient 
 is given by\cite{lor95,lor95b},
\begin{equation}
\alpha(\omega) =\alpha _0\omega I(\omega -\omega _{\parallel })\,,  \label{sdw}
\end{equation}
where $\alpha _0$ determines the oscillator strength and is given by

\begin{equation}
\alpha _0=\frac{4\pi ^2q_{{\rm A}}^2}{c\sqrt{\epsilon _1}MV_{{\rm Cu}}\omega
_{\parallel }}\,,  \label{si0}
\end{equation}
$M$ is the O mass, $V_{{\rm Cu}}$ is the volume associated with a Cu ion, $%
\omega _{\parallel }$ is the frequency of the Cu-O stretching mode phonons, $%
\epsilon _1$ is the real part of the dielectric constant, $q_{{\rm A}}$ is
an effective charge associated with this process and is of order $|q_{{\rm A}%
}|\sim e\beta aJ/(S\Delta ^2)$, where $\beta $ is the rate of change of the
Cu on-site energy due to a nearby O-ion displacement, $a$ is the lattice
constant, $S=1/2$ is the magnitude of the spin and $\Delta $ the on-site
energy difference between Cu and O.

The function $I$ is given by

\begin{equation}
I(\omega )=-16\sum_p\sin ^4(p/2)%
\mathop{\rm Im}
G(p,\omega ).
\label{idw}
\end{equation}
$I(\omega )$ has a logarithmic singularity at $\omega =J\pi /2$ which will
be convoyed to the optical absorption. The position of the singularity is a
very robust feature which comes from the saddle point at $p=\pi /2$ in the 
{\it exact} des Cloizeaux-Pearson\cite{clo62} dispersion relation (See Fig. 
\ref{mdwp}). The distance between $\omega _{\parallel }$ and the singularity
in the data provides a quick way to obtain the microscopic parameter $J$
directly from experiments.

The two-dimensional version of this theory\cite{lor95,lor95b} within an
interacting spin-wave approach was recently applied to explain related
mid-IR bands measured by Perkins{\it \ et al.} in spin 1/2\cite{per94} and
spin 1\cite{per95} insulating materials. Unlike estimations done in higher
dimension\cite{lor95,lor95b,sin89} the $J$ obtained from the position of the
singularity in 1D is ``exact'' since the location of the singularity is known
exactly. Here the only approximation is the Heisenberg model itself.

\begin{figure}[tbp]
\epsfxsize=8cm
$$
\epsfbox[50 419 562 743]{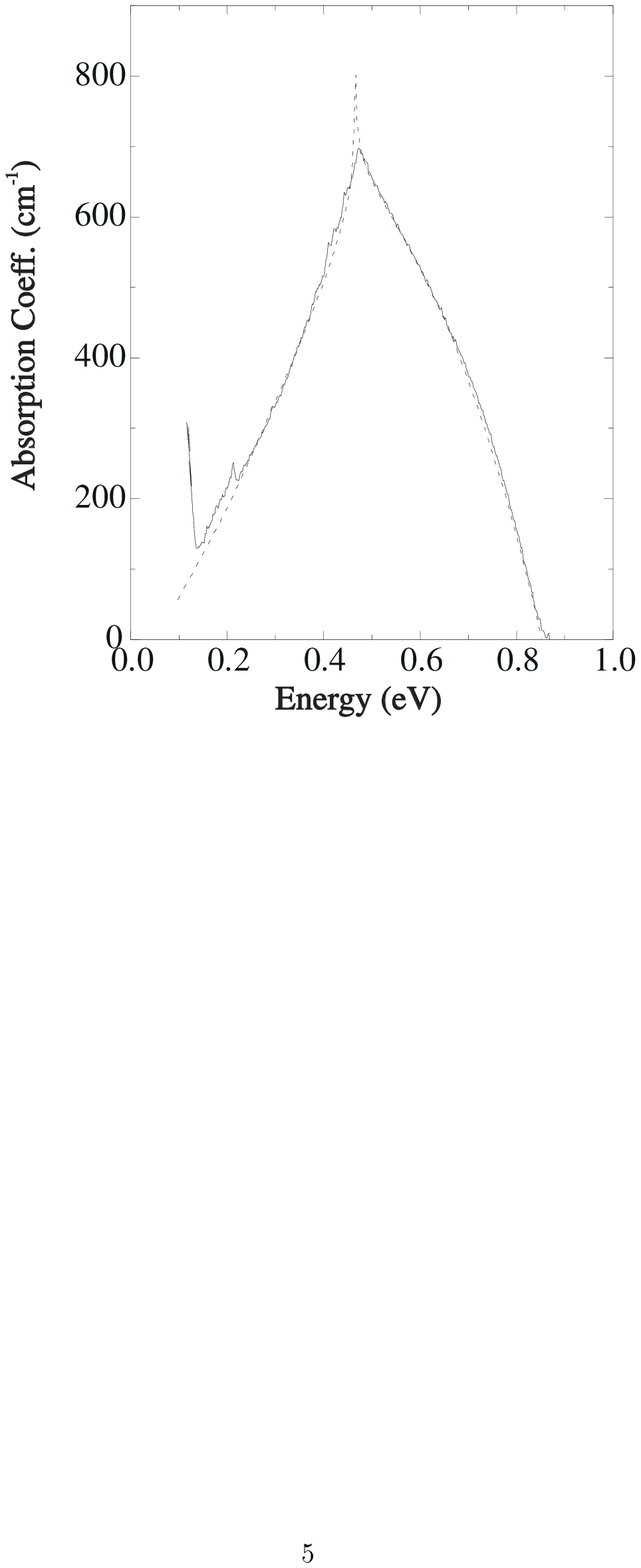}
$$
\caption{Experimental absorption coefficient from Ref.~\protect\cite{tok96}
(solid line) and theoretical line shape (dashed line) in Sr$_2$CuO$_3.$
Parameters are $J=0.246$ eV, $\omega _{\parallel }=0.08$ eV and 
$\alpha_0=132$cm$^{-1}$ . $\omega _{\parallel }$ was taken similar to 
the value in the 2D
materials. A linear background was subtracted from the experimental data.}
\label{adw}
\end{figure}

In Fig. \ref{adw} we compare the calculated  optical absorption line
shape to recent optical data by  Tokura and collaborators\cite{tok96,suz96}
in Sr$_2$CuO$_3$.
The agreement is excellent. We mention that the agreement, specially close
to the singularity is not so good if $B$ is set to zero. This suggest
(in accordance  with the numerical evidence)  that
the exponent predicted by bosonization at low energies persist up to
energies quite far from its range of applicability. 

$J=0.246$eV was 
adjusted and found to compare very well with the value obtained in
susceptibility measurements%
\cite{ami95}. 

The value of $\alpha _0$ which was fitted to the observed
spectral weight, can be used to estimate the effective charge $q_{{\rm A}}$.
We find $|q_{{\rm A}}|\simeq 0.4e$ whereas in the 2D cuprates $|q_{{\rm A}|%
}\simeq 0.1e$ was found. This is not surprising since the processes that give
rise to $q_{{\rm A}}$ are quite similar to the ones that give rise to $J$
and the latter is found to be a factor of 2 larger in the 1D compounds.

Both the singularity and the cutoff at $\omega _2(\pi )+$ $\omega
_{\parallel }$ reflect the dominance of the spectra by two-spinon states
analogous to the two-spinon states considered in Ref. \cite{bou96} for the
dynamical structure factor.

In conclusion we presented a simple ansatz for the energy dynamical
correlation function of the Heisenberg model. The ansatz reproduces very well
numerical data and the low energy behavior predicted by bosonization studies
and provides new insight on the dynamical properties of the 1D Heisenberg

After this work was completed we became aware of Ref. \cite{suz96} which
contains the experimental data and a theoretical analysis of the line shape 
{\it qualitatively} similar to ours. They predicted the same position for
the singularity however their matrix elements where computed in the XY
limit. Although at first glance the theoretical line shape resembles the
experiment they have missed the factor $\omega $ in Eq. (\ref{sdw})
and the $p$-dependent form factor in Eq.~(\ref{idw}). When 
they are included the resemblance is partially lost. 
Additionally they get a cusp singularity instead of a logarithmic singularity.

We thank Prof. Tokura for sending us their data previous to
publication, and George Sawatzky and Hamid Bougourzi for enlightening
discussions, and Y. Otha for help with the exact diagonalization program. J.
L. thanks the Lab. L\'{e}on Brillouin (CE de Saclay) for hospitality,
a postdoctoral fellowship granted by the Commission of the European
Communities at the early stages of this work and the I.S.I. foundation for
hospitality through the EU Contract ERBCHRX-CT920020. This investigation was
partially supported by Consejo Nacional de Investigaciones Cient\'{\i}ficas
y Tecnicas (CONICET), the Netherlandese Foundation for Fundamental Research
on Matter (FOM) with financial support from the Netherlands Organization for
the Advance of Pure Research (NWO).


\begin{thebibliography}{10}

\bibitem{mul79}
G. Muller, H. Beck, and J. Bonner, Phys.\ Rev.\ Lett. {\bf 43},  75  (1979).

\bibitem{mul81}
G. {Muller {\it et al.}}, Phys.\ Rev.\ B {\bf 24},  1429  (1981).

\bibitem{bou96}
A. Bougourzi, M. Couture, and M. Kacir, preprint q-alg/9604019
(unpublished); M. Karbach G. M\"uller and A. H. Bougourzi, preprint 
cond-mat/9606068 (unpublished).

\bibitem{hal82}
F.~D.~M. Haldane, Phys.\ Rev.\ B {\bf 25},  4925  (1982).

\bibitem{hal82a}
F.~D.~M. Haldane, Phys.\ Rev.\ B {\bf 26},  5257  (1982).

\bibitem{cro79}
M.~C. Cross and D.~S. Fisher, Phys.\ Rev.\ B {\bf 19},  402  (1979).

\bibitem{lor95}
J. Lorenzana and G.~A. Sawatzky, Phys.\ Rev.\ Lett. {\bf 74},  1867  (1995).

\bibitem{lor95b}
J. Lorenzana and G.~A. Sawatzky, Phys.\ Rev.\ B {\bf 52},  9576  (1995).

\bibitem{tok96}
Y. Tokura and H. Yasuhara, private communication (unpublished).

\bibitem{suz96}
H. Suzuura, H. Yasuhara, A. Furusaki, N. Nagaosa and Y. Tokura,
"preprint" (unpublished).

\bibitem{zub60}
D.~N. Zubarev, Sov.\ Phys.\ Uspk. {\bf 3},  320  (1960), [Usp. Fiz. Nauk {\bf
  71}, 71 (1960)].

\bibitem{sol79}
S{\'o}lyom, Adv. in Phys. {\bf 28},  261  (1979).

\bibitem{clo62}
J. des Cloizeaux and J.~J. Pearson, Phys.\ Rev. {\bf 128},  2131  (1962).

\bibitem{per94}
J.~D. Perkins, Ph.D. thesis, Massachusetts Institute of Technology, Cambridge,
  Massachusetts, 1994.

\bibitem{per95}
J.~D. Perkins {\it et~al.}, Phys.\ Rev.\ B {\bf 52},  R9863  (1995).

\bibitem{sin89}
R.~R.~P. Singh, P.~A. Fleury, K.~B. Lyons, and P.~E. Sulewsky, Phys.\ Rev.\
  Lett. {\bf 62},  2736  (1989).

\bibitem{ami95}
T. {Ami {\it et al.}}, Phys.\ Rev.\ B {\bf 51},  5994  (1995).

\end{thebibliography}

\end{document}